\documentstyle[12pt]{article}
\begin{document}

\title{{\bf Standing Spin Wave Resonances in Manganite Films}\\
}
\author{}
\date{}
\maketitle

\begin{flushleft}

S. E. Lofland*,S. M. Bhagat*, C. Kwon{\dag}, M. C. Robson{\dag},
 R. Ramesh{\dag}{\ddag} and R. P. Sharma{\dag}\\

*Department of Physics, University of Maryland, College Park, Maryland
20742-4111\\

{\dag}Center for Superconductivity Research, University of Maryland, College
Park, Maryland 20742-4111\\

{\ddag}Department of Materials Research and Nuclear Engineering, University
of Maryland,
 College Park, Maryland 20742-4111\\
\end{flushleft}

\newpage

Doped manganites, such as La$_{0.66}$Ba$_{0.33}$Mn0$_3$, have recently
evoked great interest because they exhibit a giant magnetoresistance [1]. It
is also  clear that the electronic transport in these materials is strongly
influenced by the spin ordering consequent upon the transition to the
ferromagnetic state. Thus, it is of fundamental interest [2] to establish
the characteristics of the low-lying excitations in the spin system.
Provided that one can obtain a suitable thin film sample, the  spin wave
resonance technique furnishes the most direct method for accessing the spin
wave stiffness $D$ in the dispersion relation, $\varepsilon = Dq^2$. In this
note we report the first observation of several well-resolved  modes in ~
100 nm thick films of La$_{0.66}$Ba$_{0.33}$Mn0$_3$. Since the modes follow
the quadratic dependence envisaged by Kittel [3], this provides a
straightforward determination of $D$.\\

Films of La$_{0.66}$Ba$_{0.33}$Mn0$_3$ (LBMO) were prepared by a
pulsed-laser deposition technique [4]. For successful observation of
standing spin wave resonances (SWR) over a wide temperature range, it was
necessary to use carefully controlled deposition parameters. Full details
will be presented elsewhere. In brief, starting with a (100) single crystal
LaAlO$_3$ substrate, we first put down a 160 nm thick SrTiO$_3$ buffer
layer, followed it with the LBMO film and eventually deposited a SrTiO$_3$
cap layer . This multilayer deposition was performed {\it in situ} using a
multi-target holder, the deposition temperature being 700$^{\circ}$ C. The
oxygen pressure was kept at 400 mTorr during deposition, and the sample was
cooled to room temperature in 300 Torr of oxygen.\\

The thickness of the LBMO film was estimated to be 110 nm using Rutherford
backscattering with a 1.5 MeV He beam obtained from a 1.7 MV tandem
accelerator. A computer fitting program was used to iteratively adjust the
thickness until the theoretical curve matched the experimental plot. The
accuracy is about 10\%.\\

Spin wave resonance (SWR) measurements were perfomed at 10 and 36 GHz for
temperatures ranging from 100 to 300 K using conventional cavity techniques
and field modulation [5].\\

Below 300 K, we observed 3 or more narrow (${\leq}$ 100 Oe) lines. For
example, Fig. 1 shows the observed spectrum at 10 GHz and 228 K, and one can
clearly discern 5 modes. The effective wavelengths lie between about 60 $a$
and 550 $a$ where $a$ is the lattice parameter; i.e, the excitations are
very close to the zone center. For a uniform ferromagnetic film whose spins
are pinned at the film surfaces, simple SWR theory [3] predicts that in the
perpendicular geometry used here, resonances occur at fields $H_n$ such that
\begin{equation}
H_n=\frac{\omega}{\gamma} +4\pi M-\frac D{\gamma \hbar}\left(
\frac n{\pi L}\right) ^2  \label{eq1}
\end{equation}

where $\omega$ is the angular frequency, $\gamma$ the gyromagnetic ratio, $M$
the magnetization, $n$ the mode number and $L$ the film thickness. Shown in
Fig. 2 is a plot of $H_n$ vs. $n^2$. The agreement with Eq. (1) is very
good. As predicted by Kittel [3] the line intensities vary roughly as $1/n^2$%
. As a further check, it was confirmed that, at a fixed temperature,
($H_{n-2} -
H_n$) is the same for both 10 and 36 GHz. In Fig. 3, we display the
normalized temperature dependence of $D$. Spin-wave theory [6] suggests
that
\begin{equation}
D(T)/D(0)=\left( 1-\alpha T^{5/2}\right)   \label{eq2}
\end{equation}

The full line in Fig. 3 represents Eq. (2) with $\alpha$ = 2.9 $\pm$ 0.3
$\times$ 10$^{-7}$
K$^{-5/2}$, a reasonable value [7]. From the data we compute that the
zero-kelvin
value is $D(0)$ = 47 $\pm$ 8 meV{\AA}$_2$. This result is at least a factor
of seven
smaller than the value estimated by Millis {\it et al}. [2] from the
measurements
of the zone-boundary magnon frequency (See Ref. 8 of [2]) in La-Pb
manganite. The magnon energy is also much lower than would be implied by the
band-theory results quoted by Millis {\it et al}.[2] Since the present
experiments
measure the magnon energies at the zone center, they are clearly more
reliable to fix the value of $D$. The only source of error in $D$ arises from
the determination of the film thickness which, as noted above, is accurate
to 10\%. Finally, one should note (Fig. 1) that the lines broaden as the mode
number is increased. A thickness variation of ${\Delta L}$ will give an
additional contribution to the linewidth $\Delta H_n$ given by
\begin{equation}
\Delta H_n=\frac{2D}{\gamma {\hbar}}\left( \frac n{\pi
L}\right) ^2\frac{\Delta L}L  \label{eq3}
\end{equation}

A 3\% variation in the film thickness is sufficient to account for the line
broadening seen in Fig. 1. It also accounts for our inability to observe
higher order resonances. In conclusion, we report measurement of spin wave
resonances in a thin film of the GMR manganite
La$_{0.66}$Ba$_{0.33}$MnO$_3$.
The zero-kelvin stiffness coefficient $D(0)$ is rather small, 47 $\pm$ 8
meV{\AA}$^2$. The $T$ dependence of $D$ is consistent with simple
spin-wave theory. It is expected that with an accurate value of $D$ becoming
available, it will be possible to achieve a more meaningful development of
theoretical models.\\

\noindent ACKNOWLEDGEMENTS. We thank the Office of
Naval Research for partial support.
\newpage
\begin{flushleft}

Figure Captions\\

Fig. 1. SWR spectrum at 10 GHz and 228 K. Note that the
field span is 6700 to 9950 Oe. One can clearly distinguish 5 distinct modes.\\

Fig. 2. $H_n$ vs. $n^2$ for 10 GHz and 228 K. The linearity, as expected for
Kittel modes, is quite satisfactory. Notice that the ordinate begins at 6000
Oe. \\

Fig. 3. Temperature dependence of the normalized spin wave stiffness.
The full curve represents $(1-2.9*10^{-7}$K$^{-5/2}T^{5/2})$.\\
\end{flushleft}
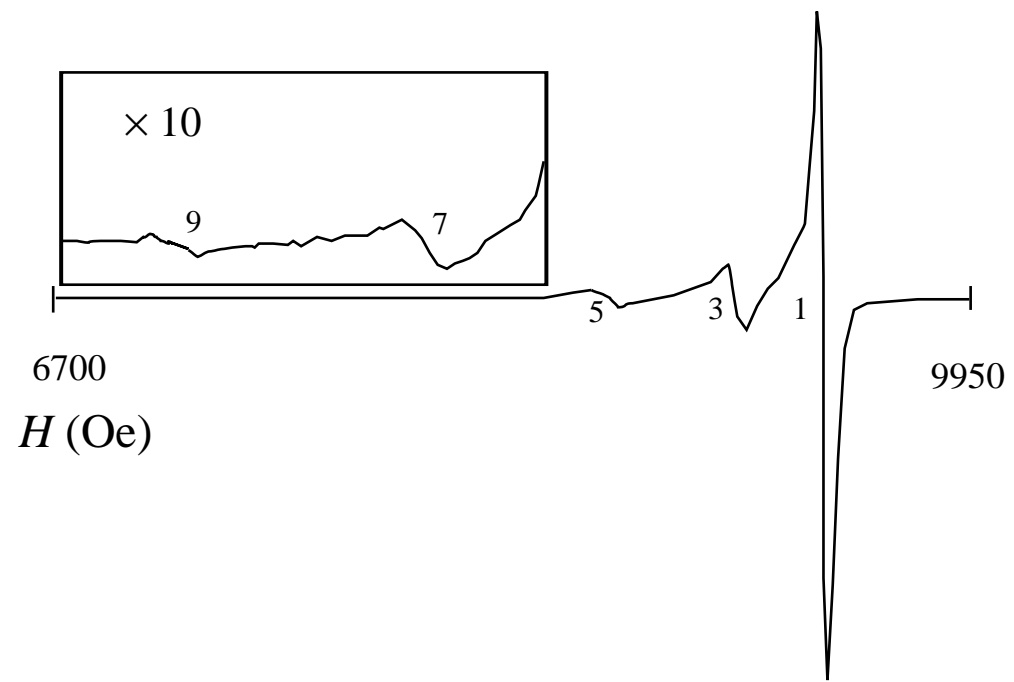
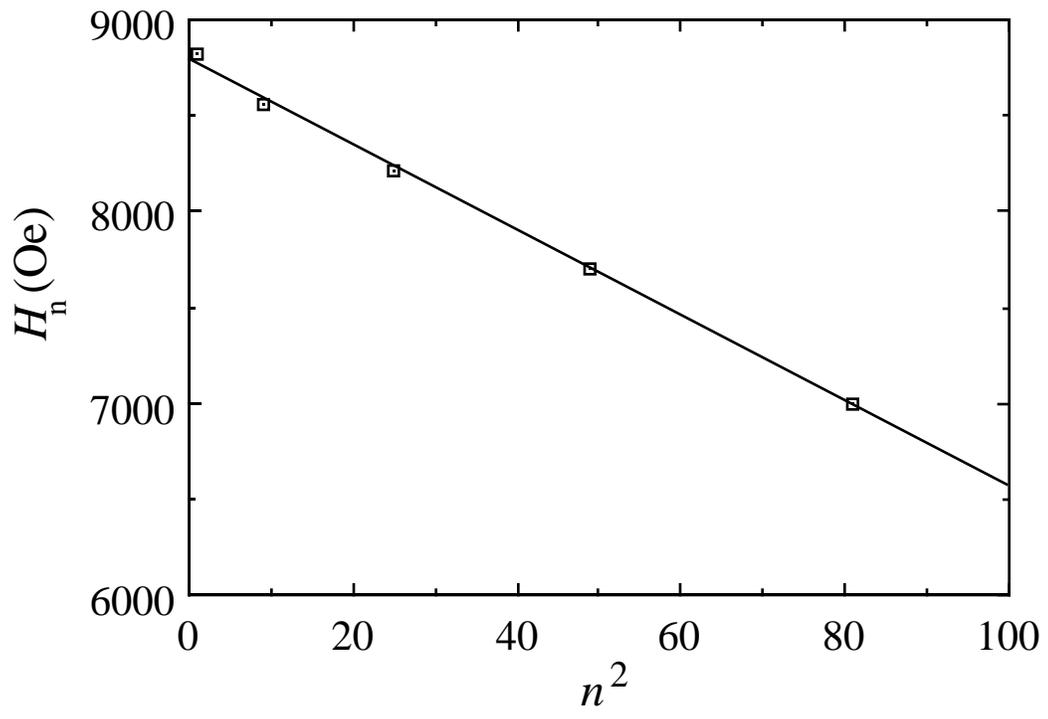
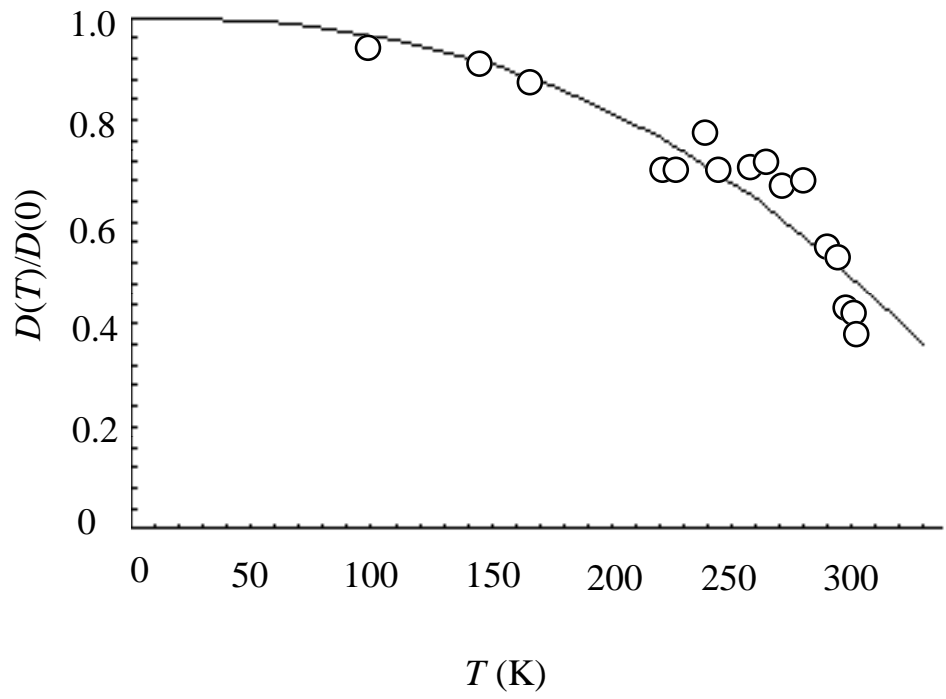
\end{document}